\documentclass[runningheads]{llncs}
\usepackage[utf8]{inputenc}
\usepackage{mwe} 
\usepackage{graphicx}
\usepackage{float}
\usepackage{hyperref}
\usepackage{multirow}
\usepackage{amsmath}
\usepackage{array}
\usepackage{tabularx}

\newcommand{\mycomment}[1]{}

\title{Assessing Generalization Capabilities of Malaria Diagnostic Models from Thin Blood Smears}
\titlerunning{Generalization Capabilities for Malaria Diagnosis}

\begin{document}

\author{Louise Guillon \inst{1} \and
Soheib Biga \inst{1} \and
Axel Puyo \inst{2} \and
Grégoire Pasquier \inst{3} \and
Valentin Foucher \inst{1} \and
Yendoubé E. Kantchire \inst{4} \and
Stéphane E. Sossou \inst{5} \and
Ameyo M. Dorkenoo \inst{4,6}  \and
Laurent Bonnardot \inst{1} \and
Marc Thellier \inst{2} \and
Laurence Lachaud \inst{3} \and
Renaud Piarroux \inst{2}
} 


\authorrunning{Guillon et al.}
%
\institute{MyC, Paris, France \\ \email{lguillon@myc.doctor} \and
APHP Paris, La Pitié-Salpêtrière, Sorbonne Université, Paris, France \and
CHU of Montpellier, University of Montpellier, Montpellier, France \and 
CHU Campus, Ministère de la Santé et de l'hygiène Publique, Lomé, Togo \and
CHU Sylvanus Olympio, Lomé, Togo \and
Faculté des Sciences de la Santé, Université de Lomé, Lomé, Togo
}

\maketitle

\begin{abstract}
Malaria remains a significant global health challenge, necessitating rapid and accurate diagnostic methods. While computer-aided diagnosis (CAD) tools utilizing deep learning have shown promise, their generalization to diverse clinical settings remains poorly assessed. This study evaluates the generalization capabilities of a CAD model for malaria diagnosis from thin blood smear images across four sites. We explore strategies to enhance generalization, including fine-tuning and incremental learning. Our results demonstrate that incorporating site-specific data significantly improves model performance, paving the way for broader clinical application.
\end{abstract}


\section{Introduction}

Malaria, caused by the parasite \textit{Plasmodium}, remains a major public health issue worldwide. According to the WHO, 249 million cases were reported in 85 countries, leading to 608,000 deaths in 2022. A fast and accurate diagnosis is crucial to avoid severe consequences. Rapid diagnosis tests have been a major breakthrough since they are easy to perform and require no special training. Yet they may be subject to false positives or false negatives and they do not provide information on the species and life stage. Therefore, microscopic analysis of thin and thick blood smears remains the gold standard for diagnosis, enabling detailed characterization of the disease. However, the diagnosis has to be made urgently and relies entirely on the expertise of the microscopist, which is variable and cannot be reviewed in real-time. Thus, automatic methods for computer-aided diagnosis (CAD) present a real opportunity and numerous works have developed such tools \cite{delahunt_automated_2015,delahunt_fully-automated_2022,yu_malaria_2020,yu_patient-level_2023,liu_aidman_2023}. Deep learning frameworks have recently achieved high performances for both thin and thick blood smear images. For thin blood smear images, accuracies of 98.62\% and 98.44\% have been reported at the cell and patient levels respectively \cite{liu_aidman_2023}. For thick smears, 91.8\%, 92.5\%, 91.1\% were obtained at the patient level for accuracy, sensitivity and specificity respectively \cite{yu_patient-level_2023}. However, most studies report results obtained at the cell level which is not clinically relevant as diagnosis is done at the patient level, relying on the results of a bench of images, each of them totalizing 200 to 400 red cells. As a matter of fact, a challenge for malaria diagnosis is to avoid false positive results at the cell scale: if only one non-infected red blood cell is predicted as positive, the whole image is wrongly predicted as positive, leading to a false positive diagnosis for the patient.

Besides, generalization is a major challenge for AI models to be used in clinical settings. Indeed, it is capital that they generalize well to all the different sites taking into account any possible site effect. As a matter of fact, various parameters can have an impact on predictions such as the quality of the microscope blades, the staining procedure, the quality and specificities of the microscope lighting, the objectives used, the smartphones and the settings used to take the photos. Additionally, parasite species distribution varies by location. Therefore, evaluating model performance across different sites is essential to ensure robustness in clinical applications. 
To our knowledge, there is no comprehensive assessment of generalization capabilities in the context of malaria diagnosis, which hampers routine use of such models. 

This work aims to evaluate the generalization capabilities of a CAD tool for malaria diagnosis and explores strategies to improve generalization. 
For that purpose, we position in a setting starting from a publicly available dataset \footnote{\url{https://lhncbc.nlm.nih.gov/LHC-downloads/downloads.html\#malaria-datasets}}. This dataset constitutes a good benchmark since it is publicly available, used by many works and collected in a clinical setting similar to ours. From this dataset, we train a baseline model and evaluate the drop of performances at cell and image levels on smartphone thin blood smear images of in-house malaria suspected cases. We then propose strategies to mitigate the site effect on our in-house datasets. Specifically, we investigate whether joint training and fine-tuning can aid in improving generalization and, if so, how many samples are needed. To the best of our knowledge, these questions have not been previously addressed.
This work is the first step in a collaboration between hospitals in non-endemic areas, industry and laboratories in endemic areas. 


Our contributions are threefold: 1) assessing the site effect on thin blood smear images, 2) evaluating generalization performance across different sites and 3) proposing joint training and  fine-tuning in an incremental learning strategy for improved model adaptability.

\section{Related works}

\subsubsection{Malaria diagnosis:}
CAD tools for malaria have been developed for several years, with recent advancements leveraging deep learning techniques. These tools can be categorized into direct detection from field-of-view images, or two-step processes, involving red cell extraction followed by classification of infected and uninfected red cells. Direct detection of infected cells can be performed thanks to object detection frameworks such as Yolo \cite{guemas_automatic_2024,yang_cascading_2020,krishnadas_classification_2022,dave_codamal_2024}, Faster-RCNN \cite{hung_applying_2017,sultani_towards_2022} or more recently based on transformers like RT-DETR \cite{guemas_automatic_2024}.
Several works have oriented their method to make it applicable to endemic zones based on specific devices like EasyScan Go \cite{delahunt_fully-automated_2022,das_field_2022} or on a setting relying on a smartphone plugged to a microscope. Thus, Malaria Screener is a smartphone app working on thin and thick smears. For thin smears, red blood cells are extracted and then classified thanks to an deep ensemble method \cite{rajaraman_performance_2019,rajaraman_performance_2019-1,rajaraman_pre-trained_2018,yu_malaria_2020,yu_patient-level_2023}. Other works have proposed solutions specific to thick blood smears to embark on smartphone apps \cite{nakasi_mobile-aware_2021}.

\subsubsection{Generalization assessment:}
Generalization is critical for clinical applications, ensuring performance consistency across diverse environments. This topic has begun to be investigated in some medical imaging applications, such as neuroimaging, histology or radiography \cite{zech_variable_2018,howard_impact_2021}. Previous studies such as those by Zech et al. on pneumonia detection, have shown performance drops when models trained on single-site data are applied to new sites, highlighting the need for diverse training data.  \cite{zech_variable_2018}. 
For malaria computer-aided diagnosis, few studies have addressed this question so far. Some works have focused on the difference in performance between several magnifications. In particular, mAP between x1000 and x400 have been reported to drop from 62.8 to 36.7 \cite{sultani_towards_2022}. To tackle this issue, domain adaptation strategies based on specific losses \cite{sultani_towards_2022} or contrastive learning \cite{dave_codamal_2024} have been proposed. Moreover, prospective studies on effectiveness of smartphone applications such as Malaria Screener \cite{yu_patient-level_2023} give some indication on their generalization. In particular, in the case of thick blood smears, the accuracy at patient-level falls from 96.7\% to 83.1\% \cite{kassim_diagnosing_2021,yu_patient-level_2023}. 


\section{Methods}
This article aims to assess the generalization capabilities of a CAD tool for malaria diagnosis. Since Yolov5 has been shown to be a relevant framework for detecting infected red blood cells and thus diagnosing malaria, we have chosen to evaluate the generalization capabilities of this model as a first step. \newline

\begin{figure}
    \centering
    \includegraphics[scale=0.15]{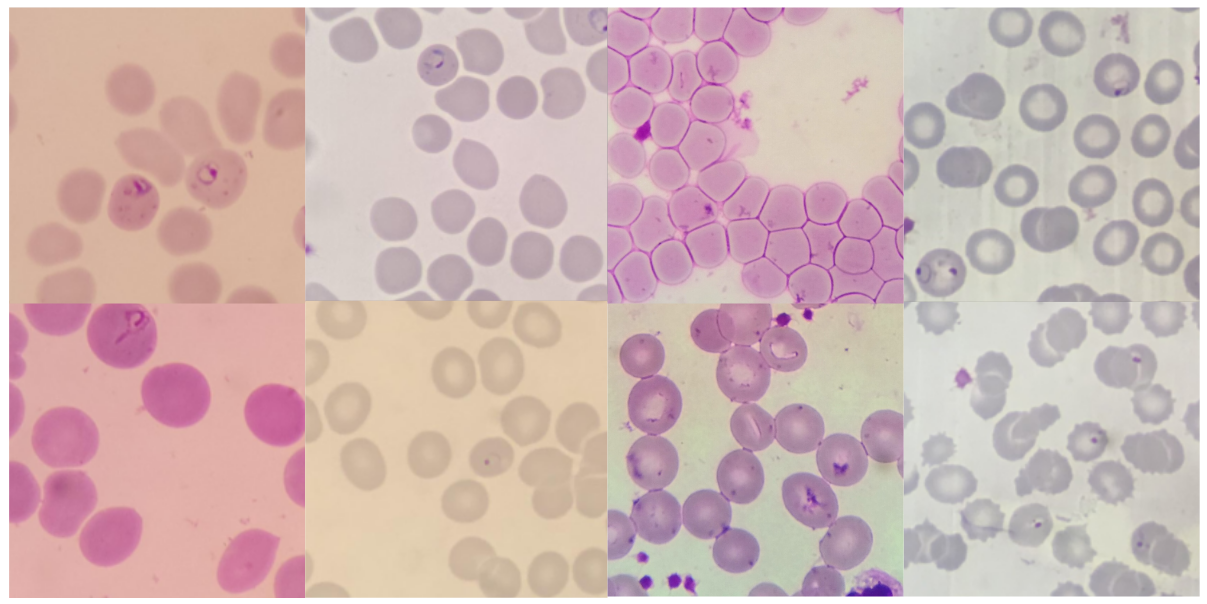}
    \caption{\textit{Zoomed cropped inputs examples.} Each column corresponds to a dataset (from left to right: NIH, hospital A, B and C). The images are zoomed crops of the input thin blood smear fields of view.}
    \label{fig:inputs}
\end{figure}

\textbf{Datasets} \newline
We use four datasets of thin blood smear images: a public dataset from the NIH \footnote{https://data.lhncbc.nlm.nih.gov/public/Malaria/NIH-NLM-ThinBloodSmearsPf/index.html} and three in-house datasets acquired in three distinct settings. All images were acquired using a smartphone connected to a microscope at x1000 magnification. All datasets but the one of the NIH contain various \textit{Plasmodium} species. For the purposes of this study, we treat all species equally as \textit{Plasmodium}. Fig. \ref{fig:inputs} shows zoomed images from the different datasets. Details are presented in table \ref{tab:data_description}.

\textbf{Training datasets.} 
Two datasets are used for the different training strategies. 
The baseline dataset is the NIH dataset. The second dataset was collected at Hospital A in a distinct country.

\textbf{Test dataset.}
We evaluate our different strategies on four datasets: a portion of the NIH dataset, a portion of the Hospital A dataset, a dataset from Hospital B and a dataset from a hospital in an endemic area (Hospital C). For the NIH and Hospital A test sets, there is no overlap with the training sets; all images of each patient are either in train or in test set.

\begin{table}
    \centering
    \footnotesize
    \begin{tabularx}{\textwidth}{|c|>{\centering\arraybackslash}X|>{\centering\arraybackslash}X|>{\centering\arraybackslash}X|>{\centering\arraybackslash}X|>{\centering\arraybackslash}X|>{\centering\arraybackslash}X|}
        \hline
        & \multicolumn{2}{|c|}{\textbf{Training data}} & \multicolumn{4}{|c|}{\textbf{Test data}} \\
        \hline
        & \textbf{NIH train} & \textbf{Hosp. A train} & \textbf{NIH test} & \textbf{Hosp. A test} & \textbf{Hosp. B} &  \textbf{Hosp. C} \\
       \hline
       Number of patients & 133 & 44 & 60 & 24 & 50 & 14   \\
       \hline
       Number of images & 665 & 671 & 300 & 311 & 200 & 59  \\
       \hline
       Number of positive images & 412 & 587 & 200 & 218 &  100 & 44  \\
       \hline
    \end{tabularx}
    \caption{Description of the datasets used. "Hosp." stands for "Hospital". "positive images" refers to images that contain at least one \textit{Plasmodium} infected red blood cell.}
    \label{tab:data_description}
\end{table}

\textbf{Malaria diagnosis prediction} \newline

To predict malaria, models are trained to detect \textit{Plasmodium} parasites in the red blood cells (RBC).
We address this as an object-detection task with infected and non-infected RBC as objects.
We study generalization on Yolo model \cite{redmon_you_2016}, which has shown high performances in detecting infected cells \cite{guemas_automatic_2024,yang_cascading_2020,krishnadas_classification_2022,dave_codamal_2024}. Specifically, we use Yolov5s pre-trained on the COCO dataset \cite{jocher_ultralyticsyolov5_2022}.
The automatic diagnosis involves two steps: first, detecting all RBC, and second, identifying infected RBC. It is necessary to detect all the red blood cells in order to count them and assess the specificity. We use two yolov5 models: model (1) detects all RBC and model (2) detects only infected RBC. Using two models enables to get rid of the class imbalance between infected and non-infected cells. To limit false positive detection, detections with a confidence threshold above 0.5 are considered positive. Detection is thus at the cell-level: a RBC is either infected or uninfected. Predictions at cell-level are then aggregated to produce a diagnosis at  image level: an image is predicted as positive if at least one RBC is infected.
\newline

\textbf{Quantification of site effect} \newline

We first quantify the site effect by assessing whether it is possible to recognize the site by just viewing an image. Using 100 positive images from NIH and Hospital A, we train a ResNet50 classifier for 20 epochs to distinguish between the two sites. To reduce the impact of different orientations and dimensions, we crop the center of each image as presented in Fig. \ref{fig:inputs}. We obtain an accuracy of $0.95 \pm 0.01$ for the Resnet50 against $0.42 \pm 0.07$ for a null model based on boostrapping. The classifier achieves a very high accuracy in predicting the site suggesting that site-specific characteristics are present in the images, potentially impacting the detection model and necessitating generalization assessment. 
\newline

\textbf{Assessing and improving generalization performances} \newline

To evaluate our framework's generalization, we train a baseline model on the NIH dataset and apply it to our other sites. To mitigate the site effect, pictures from another site can be used to update and improve the model in an incremental and finetuning strategy setting. Performances of the different strategies are compared between the 4-sites test sets.

\textbf{Baseline model.} The baseline model consists in the two yolov5 models (1) and (2) trained on the NIH data (n=665). Since model (1) is used only to compute the metrics, our mitigation strategies only concern model (2).

\textbf{Joint training model.} Our first strategy to mitigate the site effect is to complement our initial NIH training dataset (n=665) with data from hospital A (n=671) to perform joint training. We thus have 1336 images for training.

\textbf{Incremental learning models.} We investigate the impact of using varying amounts of Hospital A data ([5, 20, 50, 100, 200]) for joint training.

\textbf{Finetuning models.} 
Finally, instead of training the whole model from scratch, we evaluate finetuning the baseline model with 200 Hospital A images. We compare three settings of finetuning: fully retraining our baseline model from scratch, freezing the backbone or freezing all but the last layer of the model.

\begin{figure}[h]
    \centering
    \includegraphics[scale=0.27]{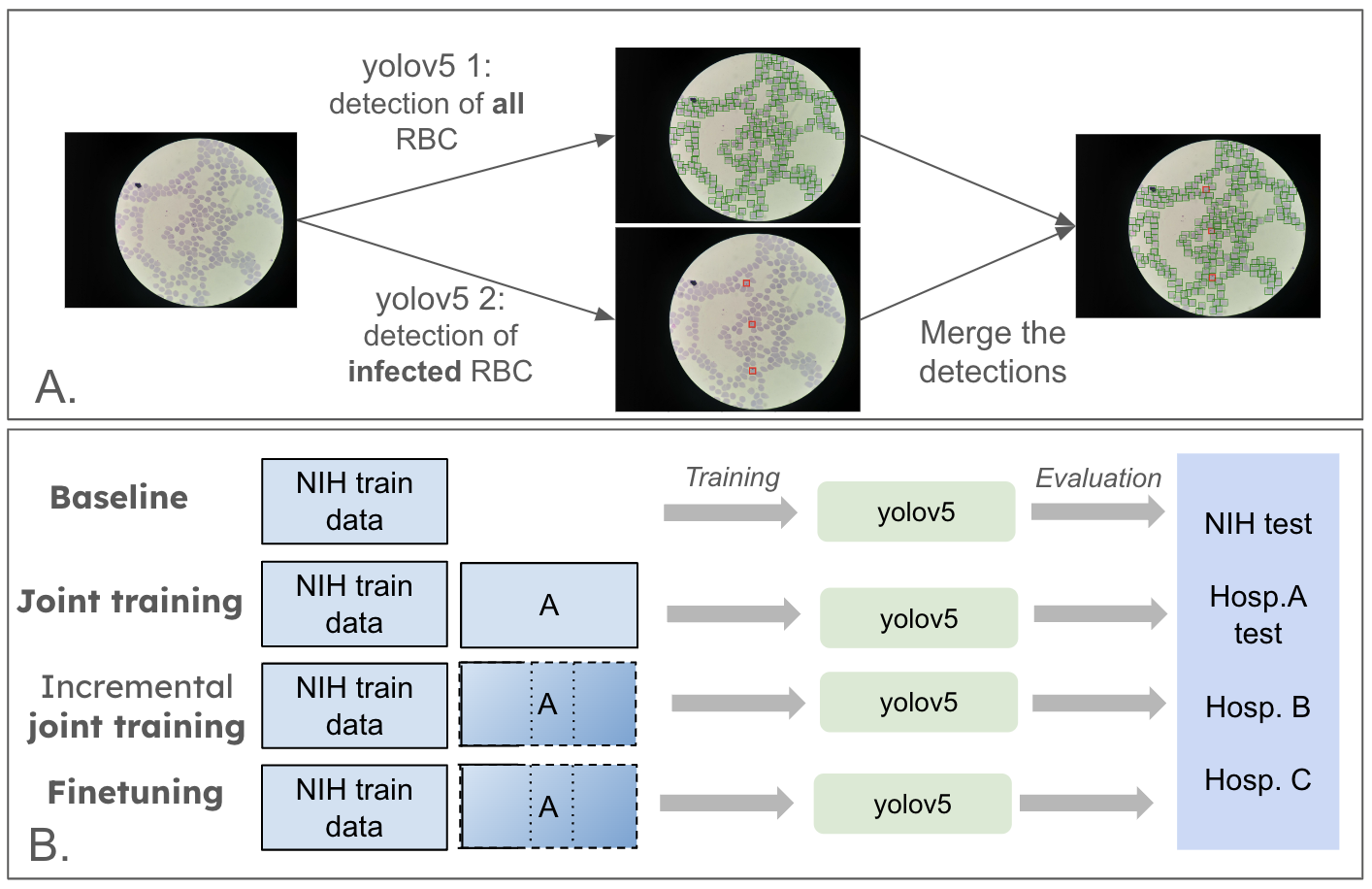}
    \caption{\textit{Framework description.} \textbf{A.} Prediction workflow. A thin blood smear field of view image is passed to two yolov5 models to detect either all RBC or only the infected RBC. Predictions are then merged to obtain final predictions of negative and positive cells. \textbf{B.} Strategies to improve generalization.}
    \label{fig:workflow}
\end{figure}

\section{Experiments and Results}
For all trainings, we performed a 5-fold cross validation, ensuring that all images from a single patient were either in the training, validation, or test set to prevent data leakage. Early stopping is applied to avoid overfitting. Training for 100 epochs takes about 2 hours on an Nvidia GeForce RTX 3050  GPU.

Table \ref{tab:test_results} presents the baseline model's performances across all the datasets.
Results on the NIH test data, with 89.2\% of accuracy at the image level and 99.08\% of accuracy at the cell level, are comparable to literature (98.62\% accuracy for Liu et al. \cite{liu_aidman_2023}and 99.61\% for Rajaraman et al. \cite{rajaraman_performance_2019}, both at the cell level).
However, as expected, there is a significant drop in performance on external datasets, with the accuracy decreasing to 97.89\% at the RBC level and 66.17\% at the image level for the Hospital A test set. This is mainly due to an increase of positive RBC predicted as negative which led to a decrease of sensibility. As a matter of fact, on an unseen dataset, model confidence is lower and our threshold of 0.5 for considering predicted positives as positives can sometimes discard true positives. 

Combining the NIH and Hospital A datasets in a joint training strategy results in an increase of the performances across most test sets. For example, the accuracy on the Hospital A test set increases significantly to 91.96\% at the image level and 98.82\% at the RBC level. Sensitivity also shows considerable improvement, indicating that the model could better identify positive cases when trained on a more diverse dataset. The improvements on hospitals B and C datasets suggest that joint training can enhance the model’s robustness to site-specific variations. For Hosp. B, although performances have improved at the RBC level, it has not led to a change of diagnosis at the image level.

\begin{figure}[h]
    \centering
    \includegraphics[scale=0.18]{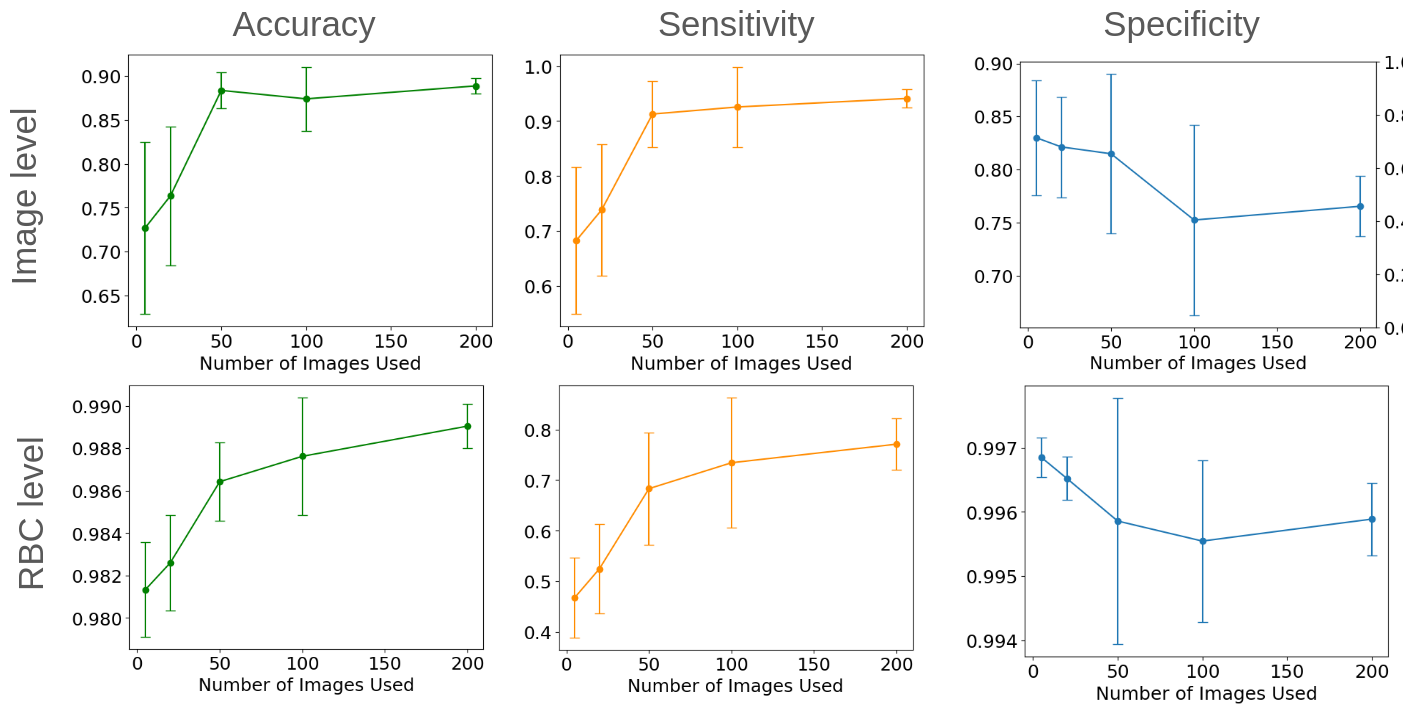}
    \caption{\textit{Incremental joint training performances.} Performance improvements observed with increasing amounts of data from Hospital A. Metrics include accuracy, sensitivity, and specificity at both image and RBC levels.}
    \label{fig:incremental}
\end{figure}

\begin{figure}
    \centering
    \includegraphics[scale=0.18]{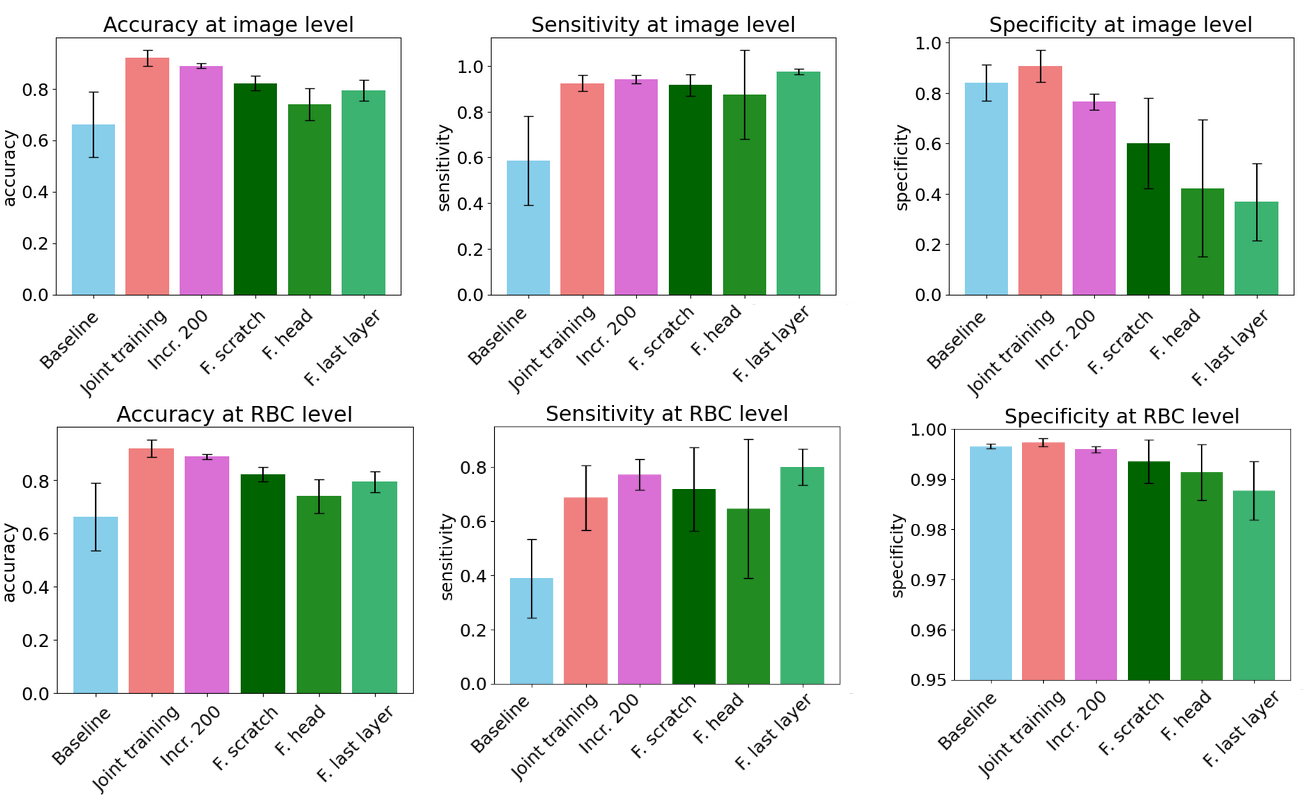}
    \caption{\textit{Finetuning performances.} Comparison of baseline, joint training with 665 and 200 images, and finetuning strategies. \textit{Incr.} and \textit{F.} stands for \textit{Incremental} and \textit{Finetuning} respectively.}
    \label{fig:finetuning}
\end{figure}

To determine the minimum amount of additional data required to improve generalization, we conducted incremental learning experiments with varying amounts of Hospital A data. Fig. \ref{fig:incremental} shows the results of joint training with increasing numbers of images from Hospital A. Even with a small number of additional samples (e.g., 20 images), there is a noticeable improvement in sensitivity at both the cell and image levels. As the number of added samples increases, the performance continues to improve, suggesting that incremental learning is an effective strategy for enhancing generalization. For 200 samples, it seems that a plateau is almost reached. This is confirmed by Fig. \ref{fig:finetuning} where there is no major difference between the joint training strategy (based on 671 Hosp. A data) and Incr. 200 which is based on 200 Hosp. A data.

The results of the different finetuning strategies using 200 images from Hospital A are shown in Figure \ref{fig:finetuning}. Finetuning the entire model or just the last layer significantly improved performance. Finetuning only the last layer was particularly effective in reducing variability, making it a robust approach for improving model generalization with minimal data. These findings suggest that finetuning is a practical method for adapting pre-trained models to new environments.

\begin{table}[h]
    \centering
    \footnotesize
    \begin{tabularx}{\textwidth}{|>{\centering\arraybackslash}X|>{\centering\arraybackslash}X|>{\centering\arraybackslash}X|>{\centering\arraybackslash}X|>{\centering\arraybackslash}X|>{\centering\arraybackslash}X|>{\centering\arraybackslash}X|}
    \hline
        \textbf{Level} & \textbf{Modality} & \textbf{Metrics} & \textbf{NIH test} & \textbf{Hosp. A test} & \textbf{Hosp. B} & \textbf{Hosp. C} \\
        \hline
        \multirow{6}{*}{Image level} & \multirow{3}{*}{Baseline} & Accuracy & $89.2_{\pm3.78}$ & $66.17_{\pm12.67}$ & $83.75_{\pm6.86}$ & $80.68_{\pm5.03}$\\
                                      &                             & Sensitivity & $88.80_{\pm7.82}$ & $58.53_{\pm19.48}$ & $83.61_{\pm7.95}$ & $91.36_{\pm3.73}$ \\
                                      &                             & Specificity & $90.00_{\pm4.64}$ & $84.09_{\pm7.11}$ & $86.67_{\pm18.26}$ & $49.33_{\pm27.73}$\\
                                      \cline{2-7}
                                      & \multirow{3}{*}{\shortstack{Joint\\training}}  & Accuracy & $90.13_{\pm1.54}$ & $91.96_{\pm3.20}$ & $83.75_{\pm8.53}$ & $84.41_{\pm8.42}$\\
                                      &                             & Sensitivity & $90.7_{\pm3.55}$ & $92.48_{\pm5.28}$ & $83.61_{\pm9.20}$ & $93.18_{\pm2.27}$\\
                                      &                             & Specificity & $89.00_{\pm3.32}$ & $90.75_{\pm6.43}$ & $86.67_{\pm18.26}$ & $58.67_{\pm38.12}$\\
        \hline
        \multirow{6}{*}{RBC level} & \multirow{3}{*}{Baseline} & Accuracy & $99.08_{\pm0.23}$ & $97.89_{\pm0.40}$ & $96.98_{\pm0.55}$ & $97.13_{\pm0.32}$\\
                                         &                              & Sensitivity & $79.58_{\pm8.33}$ & $38.81_{\pm14.57}$ & $62.07_{\pm13.53}$ & $52.42_{\pm12.33}$\\
                                         &                              & Specificity & $99.77_{\pm0.063}$& $99.65_{\pm0.048}$ & $98.92_{\pm0.60}$ & $99.26_{\pm0.25}$\\
                                         \cline{2-7}
                                         & \multirow{3}{*}{\shortstack{Joint\\training}} & Accuracy & $99.10_{\pm0.16}$ & $98.82_{\pm0.27}$ & $97.83_{\pm0.56}$ & $97.30_{\pm0.44}$\\
                                         &                              & Sensitivity & $78.55_{\pm5.86}$ & $68.62_{\pm11.90}$ & $68.55_{\pm14.29}$ & $55.08_{\pm14.81}$\\
                                         &                              & Specificity & $99.83_{\pm0.0.44}$ & $99.73_{\pm0.077}$ & $99.50_{\pm0.25}$ & $99.31_{\pm0.29}$\\
        \hline
    \end{tabularx}
    \caption{\textbf{Performances of the different strategies.} Performances are reported at image and cell levels for baseline model and the joint training strategy, with a threshold confidence at 0.5 on yolov5 (2).}
    \label{tab:test_results}
\end{table}

Overall, our results demonstrate the importance of evaluating and improving the generalization capabilities. The baseline model, while highly accurate on the training data, showed reduced performance on external datasets. Our proposed strategies enable a partial restoration of performance, suggesting practical approaches for deploying AI models in diverse clinical settings.

\section{Discussion and Conclusion}
In this work, we evaluated the generalization capabilities of a deep learning framework for malaria diagnosis from thin blood smear images. As a first step, we focused on Yolov5, a widely used model in the domain. We highlighted the impact of site-specific factors on model performance and proposed strategies to mitigate these effects. Our findings suggest that incorporating diverse data sources and employing incremental learning and finetuning techniques can significantly enhance Yolov5 generalization. 
In a clinical routine setting of an endemic area, gathering 200 images can be achieved rather quickly. Another way to improve generalization could be to perform a threshold calibration for each site, similar to the prospective study by Yu et al. \cite{yu_patient-level_2023}. 
Future works will extend this analysis to other malaria prediction frameworks such as RT-DETR \cite{guemas_automatic_2024} and explore whether specific augmentations and data preprocessing can play a role in the site effect mitigation. Finally, active learning and domain adaptation should be considered in future works, in complement to the incremental learning and finetuning strategies.

\textbf{Prospect of application.} This research will be applied through our collaboration with three hospitals and a start-up offering a cloud platform for medical facilities, currently used in clinics in malaria endemic zones. Our model, integrated into the platform, allows partners to upload smartphone photos for diagnosis. This work guides our strategy for broad deployment across diverse settings and addressing generalization capabilities.

\newpage

\bibliographystyle{splncs04}
\bibliography{paper}

\end{document}